\documentclass[onecolumn,showpacs,showkeys,preprintnumbers,amsmath,amssymb]{revtex4}

\usepackage{graphicx}
\usepackage{dcolumn}
\usepackage{bm}

\begin{document}


\title{On the boundary conditions for eliminating negative energy states in the infinite square well model}


\author{Chyi-Lung Lin}
\email{cllin@scu.edu.tw}

\affiliation{Department of Physics,\\ Soochow University, \\Taipei,Taiwan, R.O.C.}






\begin{abstract}
{ 
For a quantum confinement model, the wave function of a particle is zero outside the confined region. 
Due to this, the negative energy states are, in fact, square integrable. As negative energy states are not physical, we need to impose some boundary conditions in order to avoid these states.
For the case of the infinite square well model, we show the possible boundary conditions that avoid the existence of the negative energy states.
The well-known boundary condition requiring the wave function be continuous on both sides of the well is one of the boundary conditions that avoids negative energy states.
However, there are other types of boundary condition that can also avoid the negative energy states. This shows that there are different branch of physical systems that can be established on the infinite square well model. 
}
\end{abstract}


\pacs{03.65.-w, Quantum mechanics; 03.65.Ge, solution of wave equations: bound states}

\keywords{Infinite square well;  Time independent Schr\"odinger equation; negative energy states;  Boundary condition;}  


\maketitle


\section{\label{sec:} Introduction }

The infinite square well is a model using infinitely high potential barrier to confine particles inside a well. 
This is a basic model that appears in most textbooks.
Due to the infinite potential energy outside the well, the value of the wave functions there is  $\Psi(x)=0$, and therefore particles are confined inside the well. 
The wave functions inside the well with an energy $E$ can be solved exactly. The wave functions solved are bounded and are thus square integrable or normalizable.
However, we should note that these normalizable solutions are not restricted to positive energy states. 
In fact, the negative energy states are also square integrable. 
In general, we do not consider negative energy states, because the corresponding wave functions diverge at $|x| \to \infty $, such as the cases in the finite well model and in the unbounded potential energy systems.
There is no such problem as being divergent at infinity in the infinite square well. 
Therefore we should face the existence of negative energy states in this model.
This consideration, in fact, offers a way for properly setting the boundary conditions in order to avoid the negative energy states. 
We discuss this in detail in what follows.

The potential energy $V(x)$ of the infinite square well is described by
\begin{equation}
\label{Eq.1}
V(x)=\left\{
\begin{array}{ll}
0,&        0<x<L, \\
 \infty,&      \text{otherwise}. 
\end{array}
\right.
\end{equation}
The time independent Schr\"odinger equation  is
\begin{equation}
\label{Eq.2}
-\frac{\hbar^2}{2m} \Psi''(x)+V(x) \Psi(x)=E \Psi(x).
\end{equation} 
For positive energies, the well-known energy eigenfunctions $\Psi_n (x)$ and the eigenvalues $E_n$ of Eq.~(\ref{Eq.2}), with $V(x)$ defined in  Eq.~(\ref{Eq.1}), are as follows \cite{b.r1,b.r2,b.r3}. We have
\begin{eqnarray}
\label{Eq.3}
&&
\Psi_n (x)=\left\{
\begin{array}{ll}
\sqrt\frac{2}{L}\ \sin(k_n  x),&        0<x<L, \\
  0,&      \text{otherwise}.
\end{array}
\right.
\\
&&
\label{Eq.4}
E_n=\hbar^2  k_n^2/ 2 m,
\end{eqnarray}
where $k_n= n \pi/L$, $n=1,2,3...$.  
These solutions are obtained by imposing the boundary conditions:
%
%
%
%
\begin{eqnarray}
\label{Eq.5}
\left\{
\begin{array}{ll}
\Psi(0)=0, 
\\
\Psi(L)=0.
\end{array}
\right.
\end{eqnarray}
These boundary conditions mean that the wave function $\Psi(x)$ is continuous on both sides of the well.
In textbooks, this type of boundary condition is put in by hand, up to now, there seems no satisfiable explanation for why we  choose these boundary conditions.
%
%

%
%

In principle, the boundary conditions should be determined from  the  Schr\"odinger equation, see the discussion in Ref.\cite{b.r2}.
For instance, from Eq.~(\ref{Eq.2}), we can determine the boundary condition of $\Psi'(x)$ at $x = 0$.
By integrating Eq.~(\ref{Eq.2}) over an infinitesimally small range of x, from $-\epsilon$ to $ \epsilon$, we obtain the change of $\Psi'(x)$ at $x=0$. We have 
then
\begin{equation}
\label{Eq.6}
\Delta \Psi'(0) 
=
 \lim_{\epsilon \to 0} \Big[
 \Psi'(\epsilon)- \Psi'(-\epsilon) \Big]
=
\frac{2m}{\hbar^2}
 \lim_{\epsilon \to 0} \int_{-\epsilon}^\epsilon V(x)\Psi(x) dx .
\end{equation}
But we can not calculate the right side of Eq.~(\ref{Eq.6}), as we encounter the problem as $\infty \times 0=?$.
This means that Schr\"odinger equation does not give us a guide on how to set the boundary conditions  \cite{b.r2,b.r3}.  

The problem of setting proper boundary conditions can be solved by considering an infinite well as the limit of a finite well \cite{Seki, Rokhsar}.
%
Then the boundary condition  obtained is the requirement of the continuity of $\Psi(x)$ on both sides of the well, i.e., Eq.(\ref{Eq.5}). This is the conventional boundary condition shown on most textbooks.
%
%

There are also other approaches from different considerations that put in boundary condition that is different from the conventional one. For instance, one  may consider putting in the boundary condition by the requirement of hermiticity of the Hamiltonian defined in the interval of $x$, $[0, L]$ \citep{yshuang}.

We may consider the boundary condition problem from the following point of view. As in the infinite square well model, the Schr\"odinger equation does not give us a guide on how to set the boundary conditions.
Suppose that we don't put in boundary conditions on the two sides of the well, then what will we get?
We would obtain a continuous energy spectrum, 
and more than that, 
we would also obtain negative energy states.  
This then gives us a reason that we need to set in proper boundary conditions for avoiding the negative energy states.
%
%
%
%
%
%

\section{\label{sec:} boundary conditions for eliminating negative energy eigenstates }

We first solve the energy eigenfunctions for the cases of $E>0$, $E=0$, and $E<0$. 
For a positive energy $E>0$, we let $E=\hbar^2 k^2/2m$, the general solution of Eq.~(\ref{Eq.2}) for this energy is
\begin{equation}
\label{Eq.7}
\Psi(x)=\left\{
\begin{array}{ll}
A e^{i k x}+B e^{-i k x},&  0<x<L, \\
  0,&      \text{otherwise}.
\end{array}
\right.
\end{equation}
where $A$ and $B$ are constants. 
There is also a zero energy $E=0$ \citep{Bowen}, the corresponding wave function is:
\begin{equation}
\label{Eq.8}
\Psi(x)=\left\{
\begin{array}{ll}
Ax+B,&  0<x<L, \\
  0,&      \text{otherwise}.
\end{array}
\right.
\end{equation}
For a negative energy $E<0$, we let  $E=-\hbar^2 q^2/2m$, the general solution of Eq.~(\ref{Eq.2}) for this energy is
\begin{equation}
\label{Eq.9}
\Psi(x)=\left\{
\begin{array}{ll}
G e^{q x}
+
H e^{-q x},&  0<x<L, \\
  0,&      \text{otherwise}.
\end{array}
\right.
\end{equation}
where $G$ and $H$ are constants. From  Eq.~(\ref{Eq.9}), we see that negative energy states are square integrable, as 
$\int_{0}^{L} |\Psi(x)|^2 dx$ is obviously finite.
We thus need to impose suitable boundary conditions in order to avoid these negative energy states.
We need impose two boundary conditions to fix the values of $G$ and $H$. The boundary conditions are adjusted in such a way that $G$ and $H$ are of the value, zero.
In what follows, we consider the following different types of boundary conditions.

\subsection{\label{sec:A1}  }

If the the boundary condition we put in is:
\begin{eqnarray}
\label{Eq.A0}
\left\{
\begin{array}{ll}
\Psi(0)&=c_1, 
\\
\Psi(L)&=c_2.
\end{array}
\right.
\end{eqnarray}
%
%
We can then determine the following results from Eqs.(\ref{Eq.9})-(\ref{Eq.A0}):

\begin{eqnarray}
&& G
=\frac{c_2 e^{k L}-c_1}{e^{2kL}-1},
\label{Eq.A1}
\\
&&H
=\frac{e^{kL}\;(c_1 e^{k L}-c_2)}{e^{2kL}-1}.
\label{Eq.A2}
\end{eqnarray}
%
%
%
%
%
%
If we do not want negative energy states, we need $G=0$ and $H=0$. Setting  $G=0$ and $H=0$ in Eqs.(\ref{Eq.A1})-(\ref{Eq.A2}),  
we then obtain $c_1=0$ and $c_2=0$. 
Thus we have
\begin{eqnarray}
\label{Eq.A3}
\left\{
\begin{array}{ll}
\Psi(0)&=0, 
\\
\Psi(L)&=0.
\end{array}
\right.
\end{eqnarray}
This is just the conventional boundary conditions stated in Eq.~(\ref{Eq.5}).
Thus, we see that the conventional boundary conditions are reasonable, 
as they avoid negative energy states.
We have then given a reason for why choosing the conventional boundary condition.  
The corresponding positive energy eigenfunctions are those stated in Eq.(\ref{Eq.3}).

We also easily show that the Hamiltonian $H$ is self-adjoint that is $H^+=H$.
For $H$ to be hermitian then the boundary term
for two arbitrary eigenfunctions $f(x)$ and $g(x)$
\begin{equation}
\label{Eq.A4}
\text{B.T.} =
\Big[ 
f^*(x) g'(x)-{f^*}'(x)g(x)
\Big] \Big|_0^L
\end{equation}
should be zero.
From Eq.(\ref{Eq.3}) and (\ref{Eq.A4}),
we note that 
the boundary term is indeed zero, and hence $H $ is hermitian over the interval $[0, L]$.

\subsection{\label{sec:A2}  }

If the the boundary conditions are chosen as:
\begin{eqnarray}
\label{Eq.B0}
\left\{
\begin{array}{ll}
L \Psi'(0)&=c_1, 
\\
L \Psi'(L)&=c_2.
\end{array}
\right.
\end{eqnarray}
%
%
This then yields
\begin{eqnarray}
&& G
=\frac{c_2 e^{k L}-c_1}{k L(e^{2kL}-1)},
\label{Eq.B1}
\\
&&H
=\frac{e^{kL}\;(c_1 e^{k L}-c_2)}{k L(e^{2kL}-1)}.
\label{Eq.B2}
\end{eqnarray}
%
%
%
%
Setting  $G=0$ and $H=0$ in Eqs.(\ref{Eq.B1})-(\ref{Eq.B2}),  we obtain $c_1=0$ and $c_2=0$. 
Thus, we have
\begin{eqnarray}
\label{Eq.B3}
\left\{
\begin{array}{ll}
L \Psi'(0)&=0, 
\\
L \Psi'(L)&=0.
\end{array}
\right.
\end{eqnarray}
As $\Psi'(x)=0$ outside the well, we see that this type of boundary condition means that
 $\Psi'(x)$ is continuous on both sides of the well.
This type of boundary condition also avoids negative energy states. 

From Eqs.(\ref{Eq.7}) and  and the boundary conditions in Eq.(\ref{Eq.B3}), the corresponding positive energies and eigenfunctions are determined as follows: 
\begin{eqnarray}
\label{Eq.B4}
&&
\Psi_n (x)
=
\left\{
\begin{array}{ll}
\sqrt\frac{2}{L}\ \cos(k_n  x),&        0<x<L, \\
  0,&      \text{otherwise}.
\end{array}
\right.
\\
&&
E_n
=
\hbar^2  k_n^2/ 2 m,
\label{Eq.B5}
\end{eqnarray}
where $k_n= n \pi/L$, $n=1,2,3..$.
We note that, quite interestingly, from Eqs.(\ref{Eq.8}) there is also a  zero energy $E=0$, which corresponds to $n=0$. The results are as follows:
\begin{eqnarray}
\label{Eq.B6}
&&
\Psi (x)
=
\left\{
\begin{array}{ll}
\sqrt\frac{1}{L}, &        0<x<L, \\
  0,&      \text{otherwise}.
\end{array}
\right.
\\
&&
E
=
0.
\label{Eq.B7}
\end{eqnarray}

From Eqs.~(\ref{Eq.B4}),  Eqs.~(\ref{Eq.B6})and (\ref{Eq.A4}), we note that, the same, the Hamiltonian $H $ is hermitian over the interval $[0, L]$.

\subsection{\label{sec:A3}  }

If the the boundary condition we put in is:
\begin{eqnarray}
\label{Eq.C0}
\left\{
\begin{array}{ll}
\Psi(0)&=c_1, 
\\
L \Psi'(L)&=c_2.
\end{array}
\right.
\end{eqnarray}
%
%
We obtain:

\begin{eqnarray}
&& G
=\frac{c_1 k L+ c_2 e^{k L}}{k L(e^{2kL}+1)},
\label{Eq.C1}
\\
&&H
=\frac{e ^{k L}(k L c_1 e^{k L}-c_2)}{k L(e^{2kL}+1)}.
\label{Eq.C2}
\end{eqnarray}
%
%
Setting  $G=0$ and $H=0$ in Eqs.(\ref{Eq.C1})-(\ref{Eq.C2}),  we obtain  $c_1=0$ and $c_2=0$. 
That is
\begin{eqnarray}
\label{Eq.C3}
\left\{
\begin{array}{ll}
\Psi(0)&=0, 
\\
L \Psi'(L)&=0.
\end{array}
\right.
\end{eqnarray}

The corresponding positive energy states are determined as follows:
\begin{eqnarray}
\label{Eq.C4}
&&
\Psi_n (x)=\left\{
\begin{array}{ll}
\sqrt\frac{2}{L}\ \sin(k_n  x),&        0<x<L, \\
  0,&      \text{otherwise}.
\end{array}
\right.
\\
\label{Eq.C5}
&&
E_n=\hbar^2  k_n^2/ 2 m,
\end{eqnarray}
where $k_n= (n+1/2) \pi/L$, $n=0, 1,2,3..$. 

This is another type of boundary condition that avoids negative energy states. 
We note that  the energy is different from the conventional one in Eq.~(\ref{Eq.4}).
From Eq.~(\ref{Eq.C4})and (\ref{Eq.A4}), the Hamiltonian $H $ is hermitian over the interval $[0, L]$.

\subsection{\label{sec:A4}  }

If the the boundary condition we put in is:
\begin{eqnarray}
\label{Eq.D0}
\left\{
\begin{array}{ll}
L \Psi'(0)&=c_1, 
\\
\Psi(L)&=c_2.
\end{array}
\right.
\end{eqnarray}
%
%
We can then determine :

\begin{eqnarray}
&& G
=\frac{c_1 + c_2 k L e^{k L}}{k L(e^{2kL}+1)},
\label{Eq.D1}
\\
&&H
=\frac{e ^{k L}( c_2 k L - c_1 e^{k L})}{k L(e^{2kL}+1)}.
\label{Eq.D2}
\end{eqnarray}
%
%
%
Setting  $G=0$ and $H=0$ in Eqs.(\ref{Eq.D1})-(\ref{Eq.2}),  we obtain  $c_1=0$ and $c_2=0$. 
That is
\begin{eqnarray}
\label{Eq.D3}
\left\{
\begin{array}{ll}
L \Psi'(0)&=0, 
\\
\Psi(L)&=0.
\end{array}
\right.
\end{eqnarray}
The corresponding positive energy states are
\begin{eqnarray}
\label{Eq.D4}
&&
\Psi_n (x)=\left\{
\begin{array}{ll}
\sqrt\frac{2}{L}\ \cos(k_n  x),&        0<x<L, \\
  0,&      \text{otherwise}.
\end{array}
\right.
\\
\label{Eq.D5}
&&
E_n=\hbar^2  k_n^2/ 2 m,
\end{eqnarray}
where $k_n= (n+1/2) \pi/L$, $n=0, 1,2,3..$. 
%
%
This is another type of boundary condition that avoids negative energy states.
From Eqs.~(\ref{Eq.D4}) and (\ref{Eq.A4}), the Hamiltonian $H $ is hermitian over the interval $[0, L]$.

Other boundary condition defined at only one point, such as:
$\Psi(0)=c_1$ and $L \Psi'(0)=c2$ leads to the trivial solution: $\Psi(x)=0$. The same for the boundary conditions imposed as:
$\Psi(L)=c_1$ and $L \Psi'(L)=c2$ 

\subsection{\label{sec:A5}  }

If the the boundary condition we put in is:
\begin{eqnarray}
\label{Eq.E0}
\left\{
\begin{array}{ll}
\Psi(0)+L\Psi'(0)&=c_1, 
\\
\Psi(L)+L\Psi'(L)&=c_2.
\end{array}
\right.
\end{eqnarray}
%
%
We can then determine :

\begin{eqnarray}
&& G
=\frac{ c_2 k L e^{k L}-c_1}{(1+k L) (e^{2kL}-1)},
\label{Eq.E1}
\\
&&H
=\frac{e ^{k L}( c_2 - c_1 e^{k L})}{(kL-1)(e^{2kL}-1)}.
\label{Eq.E2}
\end{eqnarray}
%
%
%
Setting  $G=0$ and $H=0$ in Eqs.(\ref{Eq.E1})-(\ref{Eq.E2}),  we obtain $c_1=0$ and $c_2=0$. 
That is 
\begin{eqnarray}
\label{Eq.E3}
\left\{
\begin{array}{ll}
\Psi(0)+L\Psi'(0)&=0, 
\\
\Psi(L)+L\Psi'(L)&=0.
\end{array}
\right.
\end{eqnarray}
The corresponding positive energy states are
\begin{eqnarray}
\label{Eq.E4}
&&
\Psi_n (x)=
\left\{
\begin{array}{ll}
\frac{1}{\sqrt{2L}} [e^{i k_n x}+
e^{-i k_n x} (\frac{k_n L+i•}{k_n L-i•})]
,&        0<x<L, \\
  0,&      \text{otherwise}.
\end{array}
\right.
\\
\label{Eq.E5}
&&
E_n=\hbar^2  k_n^2/ 2 m,
\end{eqnarray}
where $k_n= n \pi/L$, $n=0, 1,2,3..$. 
%
%
This gives another type of boundary condition that avoids negative energy states.
From Eqs.~(\ref{Eq.E4}) and (\ref{Eq.A4}), the Hamiltonian $H $ is hermitian over the interval $[0, L]$.

\section{\label{sec:A4} conclusion  }

We have shown that although Schr\"odinger equation does not fix the boundary condition for connecting $\Psi(x)$ at the boundaries of the well; however, the requirement of not appearing the negative energy states forces us to impose some kind of boundary conditions. We have shown five types of boundary conditions that  avoid the negative energy states. 
Thus, the infinite square well model allows a branch of physical systems. Each of these systems is accompanied with a proper boundary condition that avoids negative energy states. 
There maybe other boundary conditions that can avoid the negative energy states. 

For three dimensional infinite circular well with radius a, the conventional boundary condition is $\Psi(r, \theta, \phi) =0$ at the boundary $r=a$. We can show that this is a good boundary condition, as it eliminates the negative energy states. 




\begin{acknowledgments}
The author is indebted to Prof. Young-Sea Huang for interesting discussions. The author would also like to thank Prof. Tsin-Fu Jiang and Prof. Wen-Chung Huang for much help. 
\end{acknowledgments}

\end{document}